\documentstyle[amssymb,aps,12pt]{revtex}
\begin{document}
\title{Effects of annealing and strain on La$_{1-x}$Ca$_{x}$MnO$_{3}$ thin films :
a new phase diagram in the ferromagnetic region }
\author{W. Prellier, M. Rajeswari, T. Venkatesan and R.L. Greene}
\address{Center for Superconductivity Research, Department of Physics, University of\\
Maryland, College Park, MD 20742, USA}
\maketitle

\begin{abstract}
Oriented, single phase thin films of La$_{1-x}$Ca$_{x}$MnO$_{3}$ have been
deposited onto (100)-oriented LaAlO$_{3}$ (0.1%
%TCIMACRO{\TEXTsymbol{<}}
%BeginExpansion
\mbox{$<$}%
%EndExpansion
x%
%TCIMACRO{\TEXTsymbol{<}}
%BeginExpansion
\mbox{$<$}%
%EndExpansion
0.5) substrates using the Pulsed Laser Deposition technique. While for some
compositions the physical properties (transport and magnetization) of the
as-grown films are higher than the bulk values, for other calcium contents
the optimized properties are obtained only after annealing under oxygen.
These data can be partly explained by changes in oxygen content, resulting
in cationic vacancies and thus self-doping effects - accompanying structural
changes, may be the cause of properties beyond the phase diagram.\ We
propose a new phase diagram for (La$_{1-x}$Ca$_{x})_{1-y}\square _{y}$MnO$%
_{3}$ (0.1%
%TCIMACRO{\TEXTsymbol{<}}
%BeginExpansion
\mbox{$<$}%
%EndExpansion
x%
%TCIMACRO{\TEXTsymbol{<}}
%BeginExpansion
\mbox{$<$}%
%EndExpansion
0.5) thin films.
\end{abstract}

\newpage

There has been a lot of recent interest in the properties of manganites such
as RE$_{1-x}$A$_{x}$MnO$_{3}$ (RE is a rare earth such as La, Pr, Nd and A
is an alkaline earth such as Ca, Ba or Sr) in particular due to a
spectacular decrease of electrical resistance under a magnetic field \cite
{VonHel}, the so-called colossal magnetoresistance (CMR) \cite{Jin}. Due to
the long history of work on these compounds, most of the studies have been
performed on bulk ceramic samples.\ For example, it was demonstrated that
chemical substitution on the trivalent ion sites can have an effect similar
to annealing (i.e. changing the Mn$^{3+}$/Mn$^{4+}$ ratio). Also, the
determination of the ''phase diagram'' upon alkaline earth doping has been
achieved essentially on ceramics samples \cite{Schiffer,Cheong}.

The basic phenomena of the CMR\ effect seems to be the same in bulk and thin
film samples. However, due to the strain effect of the substrate or oxygen
deficiency, it is often difficult to reach in a thin film the same
properties as the bulk \cite{Ju,Xiong,Wil}. But surprisingly, it has been
shown recently that an anomalously high metal-insulator transition ($T_{MI}$%
)\ occurs in thin films of nominal composition La$_{0.8}$Ca$_{0.2}$MnO$_{3}$ 
\cite{Shree} ($T_{MI}$ is near room temperature, 100\ K higher than the bulk
value \cite{Schiffer}). For these reasons, we decided to investigate the
effect of doping and annealing in thin films deposited on LaAlO$_{3}$
substrates for the solid solution La$_{1-x}$Ca$_{x}$MnO$_{3}$ (0%
%TCIMACRO{\TEXTsymbol{<}}
%BeginExpansion
\mbox{$<$}%
%EndExpansion
x%
%TCIMACRO{\TEXTsymbol{<}}
%BeginExpansion
\mbox{$<$}%
%EndExpansion
0.5). We have compared the properties of as-grown films and post-annealed
ones under oxygen or argon flux. We have obtained the phase diagram as a
function of temperature (T) and x, for both as-grown and oxygen annealed
films, which is significantly different from the bulk one \cite
{Schiffer,Cheong}. These data are compared and discussed with bulk ceramics.

Thin films of La$_{1-x}$Ca$_{x}$MnO$_{3}$ (LCMO) were grown using the Pulsed
Laser Deposition (PLD) technique. The targets used had a nominal composition
of La$_{1-x}$Ca$_{x}$MnO$_{3}$ (x=0.1, 0.15, 0.2, 0.25, 0.33, 0.4 and 0.5).
Most of the films were synthesized on [100] LaAlO$_{3}$ substrates, which
has a pseudocubic crystallographic structure with a=3.79 \AA . The laser
energy density on the target was about 1.5 J/cm$^{2}$ and the deposition
rate was 10 Hz. The substrates were kept at a constant temperature of 820 ${%
{}^{\circ }}$C during the deposition which was carried out at a pressure of
400 mTorr of oxygen. After deposition, the samples were slowly cooled to
room temperature at a pressure of 400 Torr of oxygen. All films had a
thickness around 1500 \AA . Further details of the target preparation and
the deposition procedure are given elsewhere \cite{Ju}. The annealing was
done under flowing oxygen or argon at 800 ${{}^{\circ }}$C for 10 h. The
heating and the cooling were done at 10 ${{}^{\circ }}$C/min. Note that the
temperature of annealing is below the synthesis temperature in order to keep
the same structure as the one obtained during the laser ablation process.
The structural study was carried out by X-ray diffraction (XRD) using a
Rigaku diffractometer with the usual $\Theta -2\Theta $ scan. DC resistivity
($\rho $) was measured by a standard four-probe method and the magnetization
(M) was obtained using a Quantum Design MPMS SQUID magnetometer.

The X-ray diffraction data indicate that all the films are single phase
since only two sharp diffraction peaks at around 2$\Theta \approx $23$%
^{\circ }$ and 46$^{\circ }$ appear (see inset of Fig.1 for x=0.25). They
correspond to an out-of-plane parameter proportional to the ideal parameter
of cubic perovskite ($\approx $3.9 \AA ). It is not possible on the basis of
this pattern to tell if the film is [110] or [001]-oriented, however, this
does not matter for the conclusions of this paper. Fig.1 shows
simultaneously the evolution of the out-of-plane parameter as a function of
x for an as-grown and an oxygen annealed film. We also show some values
given by Huang et al. \cite{Huang} on ceramic samples. Two features can be
seen from this graph. First, a small decrease of the out-of-plane parameter
with increasing calcium content is found and this is consistent with the
formation of mixed valence Mn$^{3+}$/Mn$^{4+}$ \cite{Wollan}. Second, for a
given composition, a decrease of this parameter is found after annealing
under oxygen and this change is more pronounced for lower doping of calcium
(for example at x=0.15, we have 3.93 \AA\ for the as-grown film and 3.87
\AA\ for the annealed film).

Fig.2 shows the DC\ magnetization taken under a magnetic field of 2000 Oe
for the as-grown and annealed film of x=0.25.\ The Curie temperature ($T_{c}$%
) is increased from about 240 K to 270\ K after oxygen annealing. At the
same time, the metal-insulator transition $T_{MI}$ (see inset of Fig.2)
increases from about 250 K to 281 K. The difference between $T_{c}$ and $%
T_{MI}$ is most likely a particle size effect \cite{Mahesh} which is
consistent with our observations that the full width at half maximum of the
diffraction peak decreases upon annealing (from 0.2$^{\circ }$ to 0.1$%
^{\circ }$ after annealing). The increase of the transition temperature
under annealing is not surprising since it is well known that this treatment
can often improve the physical properties of manganites and also other
oxides \cite{Chang}. However, it is surprising that both $T_{MI}$ and $T_{c}$
are much higher than the bulk values previously reported by Shriffer et al. 
\cite{Schiffer}.

The saturation magnetization value, obtained with a magnetic field of 2 T,
is 3.75 $\mu _{B}$ for the as-grown film (the expected theoretical value),
but is 3.59 $\mu _{B}$ after oxygen annealing. This change can be
interpreted as an increase of the Mn$^{4+}$/Mn$^{3+}$ ratio after annealing.
An increase of Mn$^{4+}$ would explain why $T_{c}$ and $T_{MI}$ also
increase, at least according to the bulk phase diagram\cite{Schiffer,Cheong}
(see Fig.5). This suggests that under oxygen annealing the oxygen content in
the film is increasing and we expect :

$La_{1-x}Ca_{x}MnO_{3}+\frac{\delta }{2}O_{2}\rightarrow
La_{1-x}Ca_{x}MnO_{3+\delta }=(La_{1-x}Ca_{x})_{1-y}\square
_{y}Mn_{1-z}\square _{z}O_{3}$

(where $\square $ represents cations vacancies and $\delta $ this excess of
oxygen)

Thus, excess oxygen leads to an equal number of vacancies at both of the
cation sites \cite{Dabrowski} and therefore to an increase of the Mn$^{4+}$
content. To check if these changes are related to oxygen incorporation or
thermally effects, we have annealed the as-grown film (La$_{0.75}$Ca$_{0.25}$%
MnO$_{3}$ on LaAlO$_{3}$) under argon and oxygen at the same temperature.
The resistivity of both samples is presented in Fig.3. The effect of argon
is negligible for the $T_{MI}$ and the out-of plane parameter remains
unchanged ($\approx $3.86 \AA ). Also, the temperature coefficient of
resistance ($1/RdR/dT$) is nearly the same. Thus, the effect of annealing is
from oxygen incorporation and not the temperature.

The shift towards the Mn$^{4+}$ rich region of the phase diagram under
oxygen annealing is more pronounced for a low doping of calcium as already
reported for bulk samples \cite{Huang2}. As an example, we show in Fig.4 the
effect of oxygen annealing for a film with composition La$_{0.9}$Ca$_{0.1}$%
MnO$_{3}$ deposited on LaAlO$_{3}$. The as-grown film is an insulator, as
expected from the bulk phase diagram, but after oxygen annealing displays a
metal-insulator transition at 170 K. Again, this is a result of increased Mn$%
^{4+}$ and is perfectly consistent with the bulk phase diagram since x=0.1,
is already close to the boundary between the insulator and metallic regions
(x$\approx $0.175). In addition, under oxygen annealing the change of the
out-of-plane parameter is very strong, consistent with the formation of
cation vacancies at the La and Mn sites \cite{Gupta}.

We summarize these data in Fig.5 with a plot of the film phase diagram for
the composition range x=0.1 to x=0.5. There is clearly a difference from the
bulk phase diagram (dashed line in Fig.5). Some of this difference can be
explained by 'self-doping' effects related to increase in Mn$^{4+}$ due to
cation vacancies. But this is not sufficient to explain why for the x=0.2
composition, $T_{MI}$\ is 30 K higher than the highest $T_{MI}$ found in any
bulk phase (i.e. $T_{MI}\approx $260 K for x=0.33). It is possible that the
presence of cationic vacancies also result in structural changes which be
more easily accommodated in thin film form (i.e. strain-induced by the
substrate).\ For instance, if the cationic vacancies would lead to an
internal pressure, i.e. decreased with cell volume, we may expect the $%
T_{MI} $ and $T_{c}$ to be higher as for hydrostatic pressure effects in
bulk \cite{Neumeier,Hwang}. We are currently undertaken detailed structural
studies.

In summary, we have investigated the structural, magnetic and transport
properties of a series of La$_{1-x}$Ca$_{x}$MnO$_{3}$ deposited on LaAlO$%
_{3} $ in order to better understand the effect of composition on the
properties of thin films. We found that the thin film phase diagram is
somewhat different than that of the bulk. Oxygen annealing leads to an
enhancement of $T_{MI}$ and $T_{c}$. We explain this as resulting from two
related contributions : the strains induced by the substrate and the change
in oxygen stoichiometry (leading to cation vacancies) due the annealing.

\medskip \centerline{\bf Acknowledgments}

Partial support of NSF-MRSEC at University of Maryland is acknowledged (DMR\
\#96-32521). We thank Z. Li for the preparation of targets, M.\ Lewis for
transport measurements and R.C. Srivastava for XRD\ measurements.. We
acknowledge A. Biswas and P.A. Salvador for helpful discussions.

\centerline{\bf Figure captions}

\begin{center}
\noindent {Fig.1 : Evolution of the out-of-plane parameter vs x in thin
films of La}$_{{1-x}}${Ca}$_{{x}}${MnO}$_{{3}}${\ deposited on LaAlO}$_{{3}}$%
{\ substrates. Solid and open squares correspond respectively to the
as-grown and annealed films. We also show the bulk lattice parameters from
Ref. [9] (open triangles : c axis, solid triangles : a,b axes). Lines are
only a guide for the eyes. The inset shows the diffractogram of as-grown La}$%
_{{0.75}}${Ca}$_{{0.25}}${MnO}$_{{3}}${\ on LaAlO}$_{{3}}${.}

\noindent {Fig.2 : DC magnetization vs temperature for a La}$_{{0.75}}${Ca}$%
_{{0.25}}${MnO}$_{{3}}${\ thin film on LaAlO}$_{{3}}${\ taken with a
magnetic field of 2000 Oe}. {Squares and circles correspond respectively to
the as-grown and the oxygen annealed films. The inset shows }${\rho }${(T)
for the same films.}

Fig.3 : $\rho (T)$ after different annealing treatment (as-grown, oxygen and
argon annealing) for La$_{0.75}$Ca$_{0.25}$MnO$_{3}$ on LaAlO$_{3}$.

Fig.4 :\ $\rho (T)$ for a La$_{0.9}$Ca$_{0.1}$MnO$_{3}$ film on LaAlO$_{3}$.
Comparison between an as-grown and oxygen annealed film.

Fig.5 : Phase diagram for La$_{1-x}$Ca$_{x}$MnO$_{3}$. $T_{c}$ and $T_{MI}$\
are taken from the inflection point in M(T) and ${\rho }${(T)}. Dashed line
indicates the data from Ref. {[3-4]. Gray dashed lines separate the
different regions. Solid lines are only a guide for the eyes.}
\end{center}

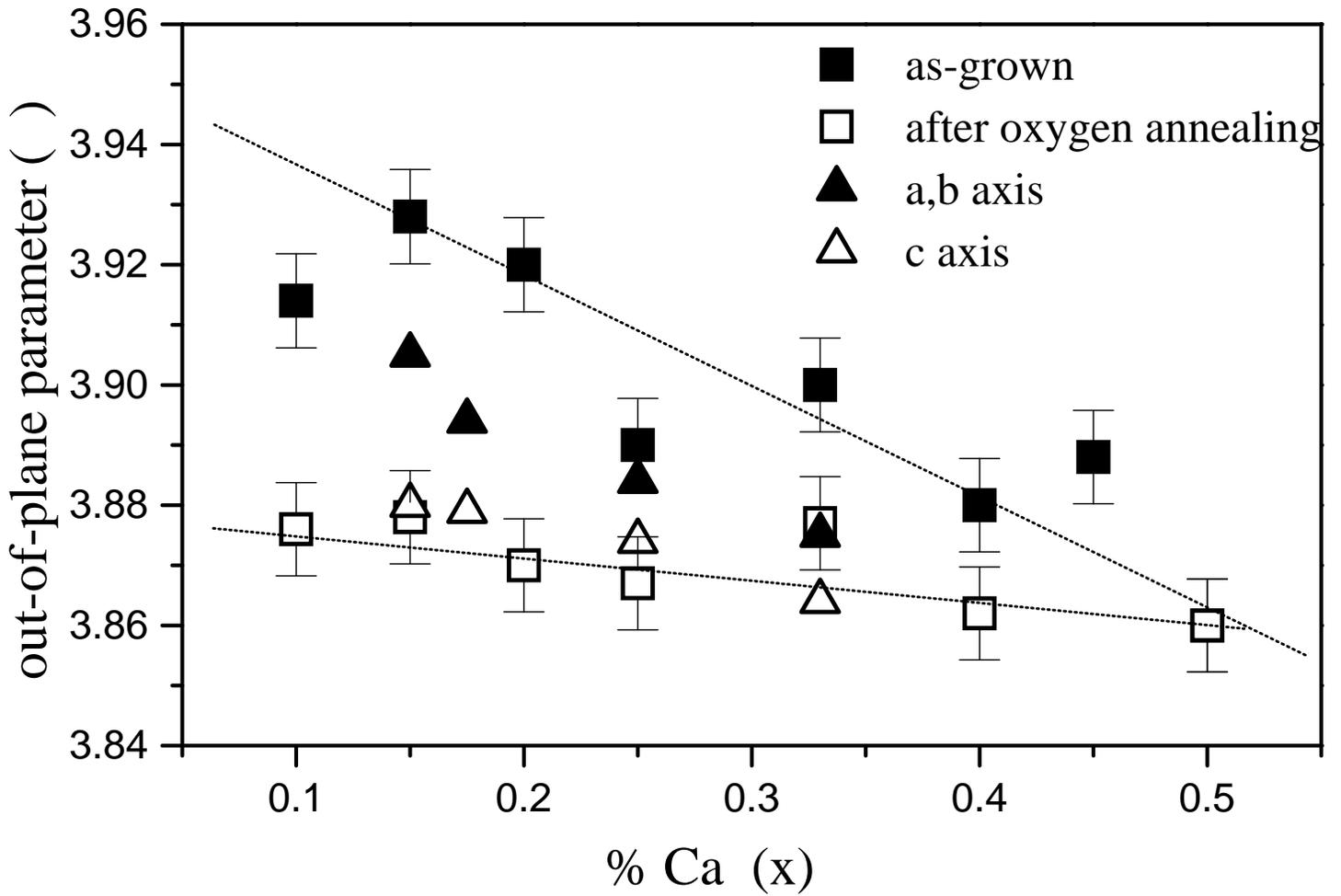
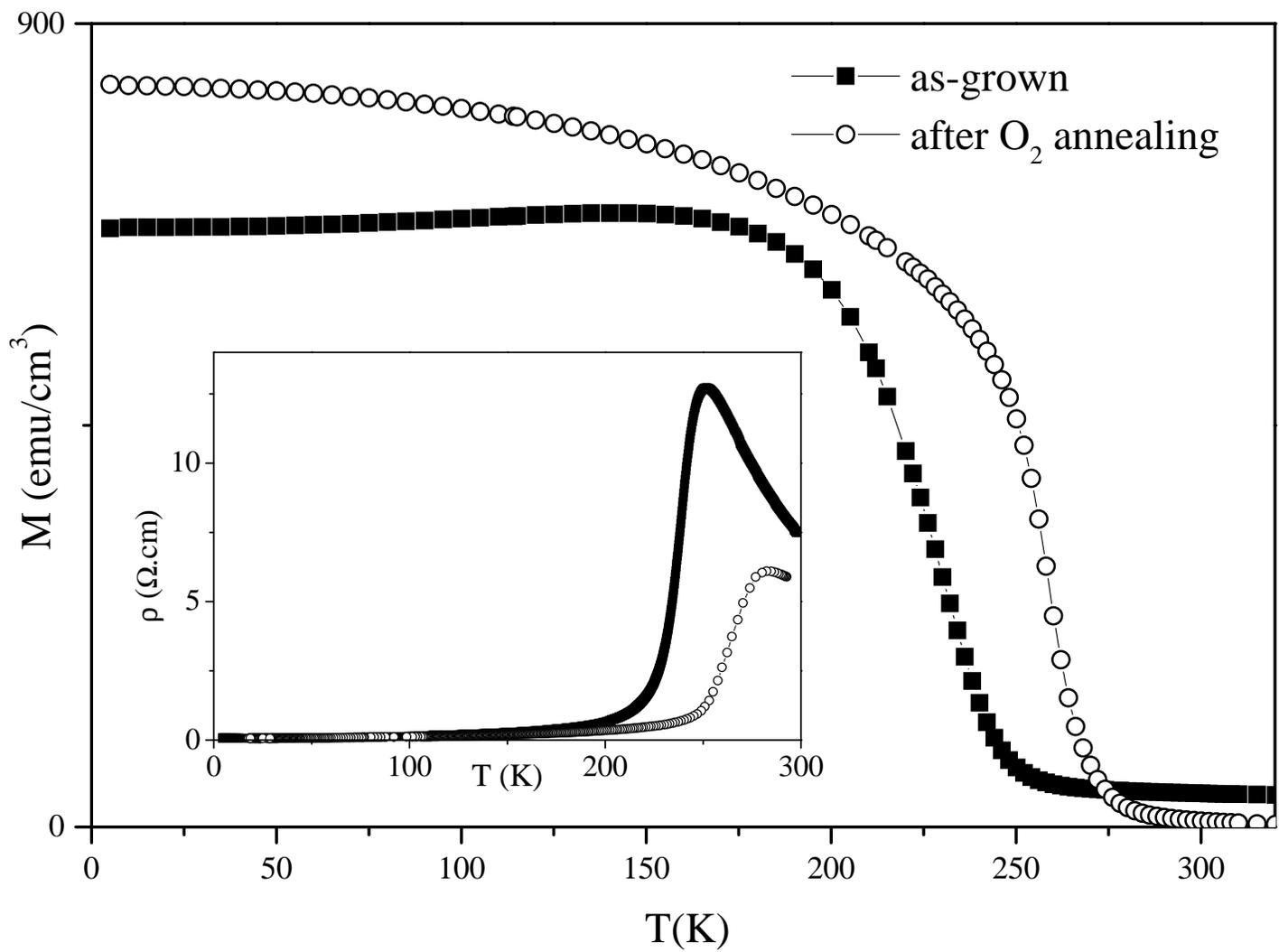
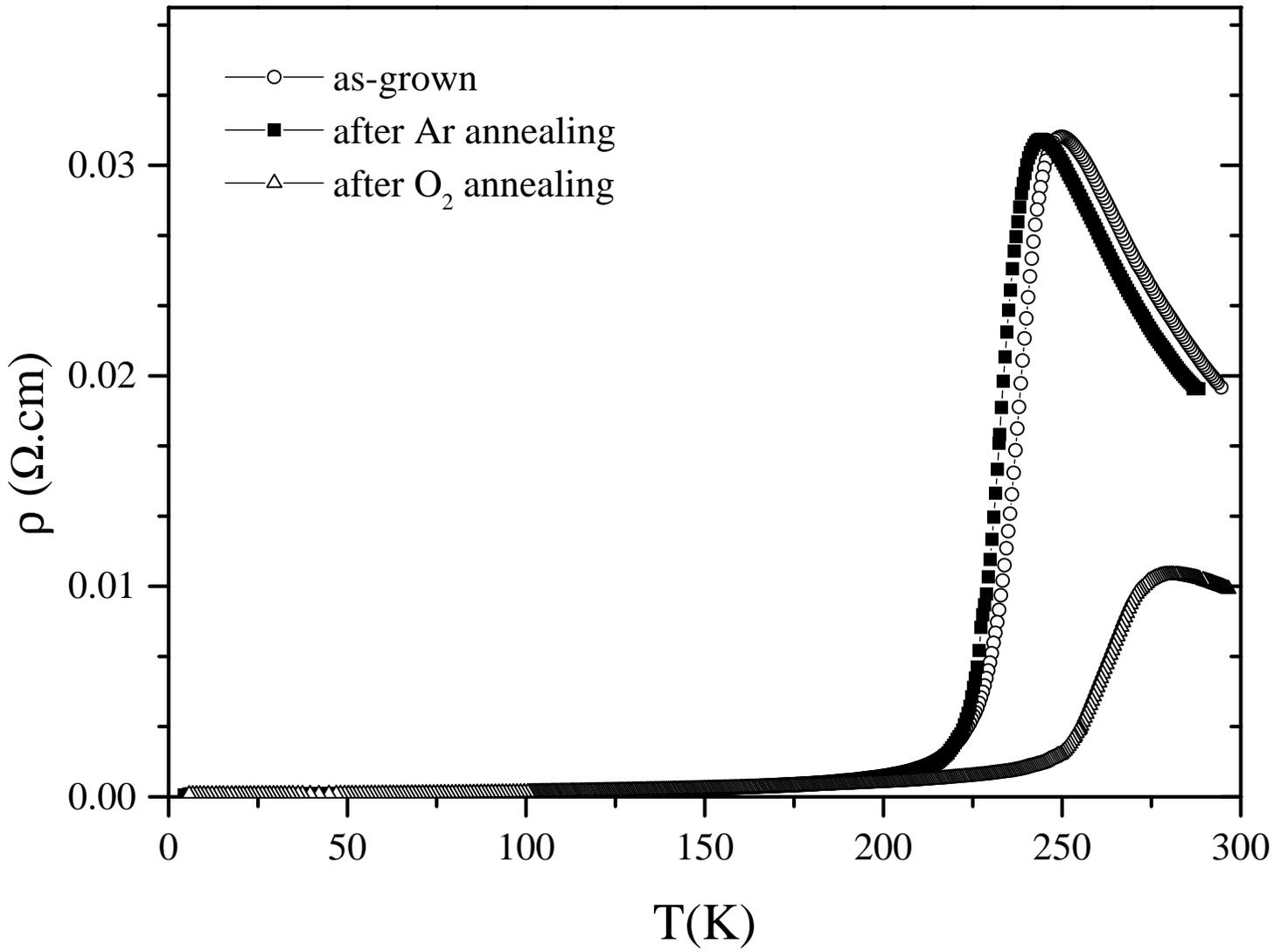
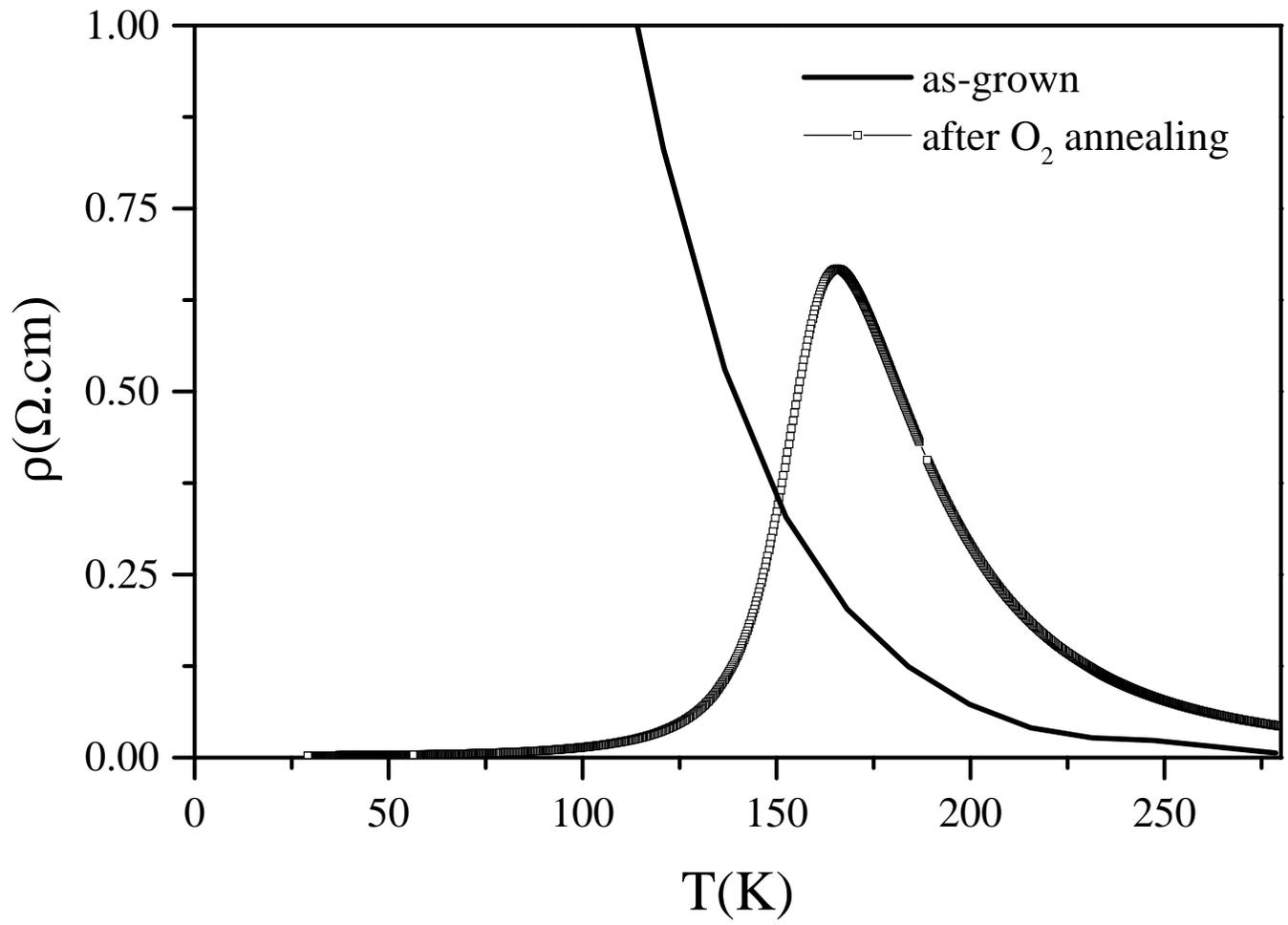
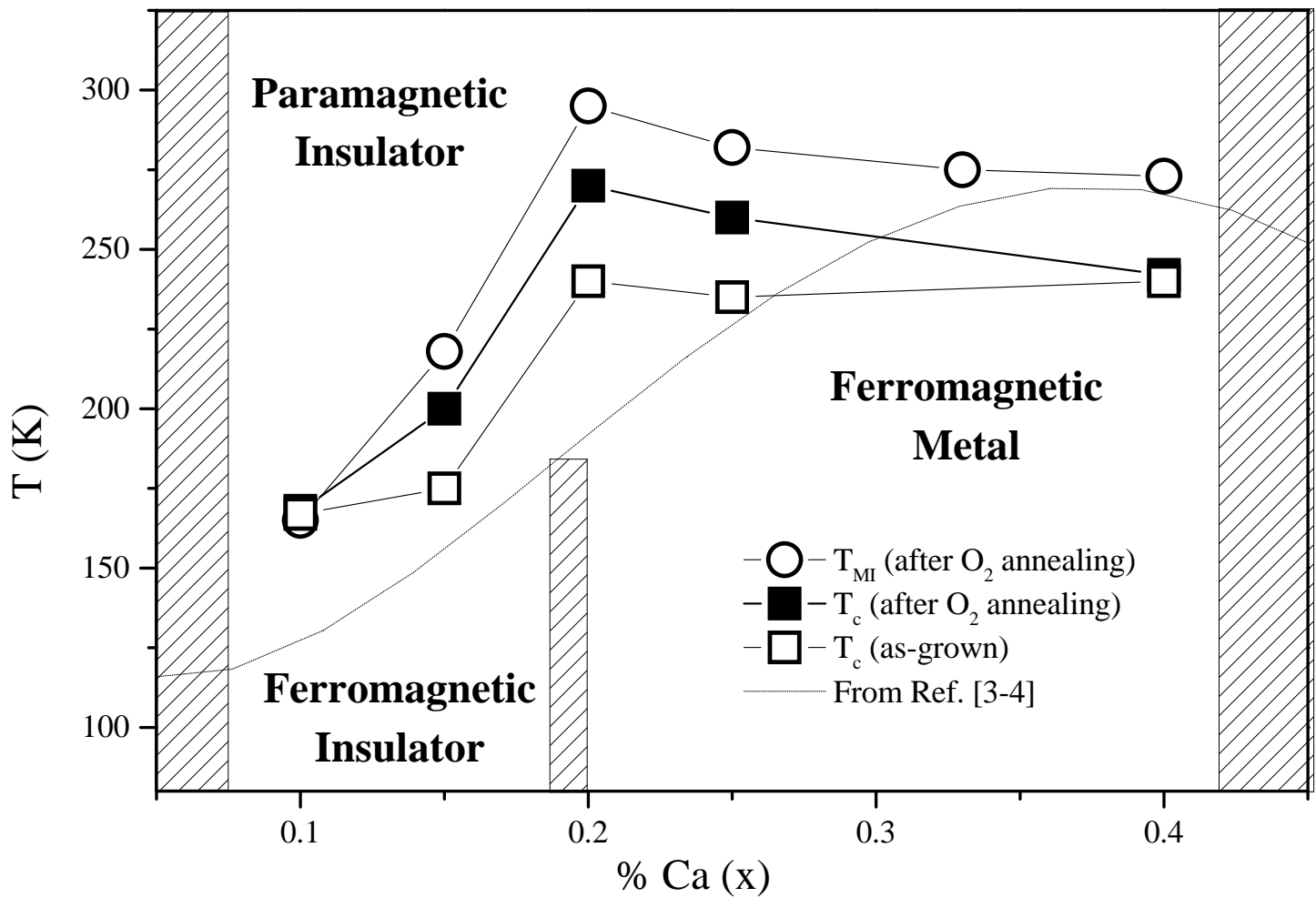
\end{document}